\documentclass[12pt,superscriptaddress,aps,prd,preprint,showpacs]{revtex4}
\usepackage[dvips]{graphicx}
\usepackage[colorlinks=true, linkcolor=blue, citecolor=blue, urlcolor=blue]{hyperref}

\usepackage{amsfonts}
\usepackage{slashed}
\usepackage{amsmath}
\usepackage{slashed}
\newcommand{\be}{\begin{equation}}
\newcommand{\en}{\end{equation}}
\newcommand{\ee}{\end{equation}}
\newcommand{\bea}{\begin{eqnarray}}
\newcommand{\eea}{\end{eqnarray}}

\def\As{A\!\!\!/}

\def\ds{\partial\!\!\!\!/}
\def\Ds{D\!\!\!\!/}
\def\Ss{T\!\!\!\!/}
\def\Ss{S\!\!\!\!/}
\def\os{\omega\!\!\!\!/}

\newcommand{\remark}[1]{}

\begin{document}

\title{Mixed gauge-gravity term and proper time}

%%%%%%%%%%%%%%%%%%%%%%%%%%%%%%%%%%%%%%%%%%%%%%%%%%%%%%%%%%
\author{ J. R. Nascimento}
\email{jroberto@fisica.ufpb.br}

\author{ M. Paganelly}
\email{matheuspaganelly@gmail.com}

\author{  A.Yu. Petrov}
\email{petrov@fisica.ufpb.br}

\author{ P.J. Porfírio}
\email{pporfirio@fisica.ufpb.br}
\affiliation{ Departamento de Física, Universidade Federal da Paraíba, Caixa Postal 5008, 58051-970, João Pessoa, Paraíba, Brazil.}

%%%%%%%%%%%%%%%%%%%%%%%%%%%%%%%%%%%%%%%%%%%%%%%%%%%%%%%%%%

%%%%%%%%%%%%%%%%%%%%%%%%%%%%%%%%%%%%%%%%%%%%%%%%%%%%%%%%%%
%%%%%%%%%%%%%%%%%%%%%%%%%%%%%%%%%%%%%%%%%%%%%%%%%%%%%%%%%%

%\preprint{}

\begin{abstract}

 We consider the four-dimensional action of spinors minimally coupled to a $U(1)$ gauge field on a Riemann-Cartan background. In this theory, we integrate over the spinors and study the resulting one-loop gauge-gravity effective action, paying special attention to the contributions that depend on both the gauge field and the torsion. We explicitly calculate the gauge-torsion term, which turns out to be finite, and comment on possible terms depending simultaneously on torsion, curvature, and gauge field.
 
\end{abstract}
\maketitle

%\vspace{1cm}

%%%%%%%%%%%%%%%%%%%%%%%%%%%%%%
\section{Introduction}

General relativity (GR) is one of the most successful theories in physics. It has been proven to be robust even after more than a century of experimental tests, such as gravitational redshift, Mercury's perihelion precession, and the deflection of light by the Sun \cite{Pireaux, Will:2014kxa, Berti:2015itd}, as well as the prediction of gravitational waves confirmed by LIGO \cite{LIGO1, LIGO2}.

However, at the end of the last century, cosmic observations revealed new intriguing physical phenomena, namely: the late-time acceleration of the Universe \cite{Garnavich,Riess, Riess2, Perlmutter, Schmidt,Steinhardt, Persic} and the discrepancy between the observed and measured quantities of the galaxies' rotation curves (known as the galaxy rotation problem). GR cannot fully explain such issues without introducing any additional component of dark exotic matter (dark energy and dark matter. In this scenario, alternative theories of gravity have provided a fruitful arena for explaining these phenomena (and others), eliminating the need for exotic matter sources.

%revealed that the Universe is undergoing accelerated expansion \cite{Garnavich}. {\color{red} }

%More recent data indicate that the total content of the Universe is predominantly composed by dark energy and dark matter, which together account for about 96\% of the total. The baryonic matter that forms known structures makes up only 4\% of the Universe \cite{Riess, Riess2, Perlmutter, Schmidt,Steinhardt, Persic}. These observations are not fully explained by General Relativity, prompting the development of extended gravitational theories.

Over time, numerous modified theories of gravity have been considered to explain new physics beyond GR. The vast majority of them propose modifications in the pure geometrical sector of GR, including considering higher-order geometrical invariants and/or non-Riemannian geometries to describe the background geometry \cite{Hehl,Olmo:2011uz}. Regarding the latter, the Einstein-Cartan theory plays a pivotal role since it was one of the first modified theories of gravity defined in a non-Riemannian background (in this case, Riemann-Cartan geometry \cite{Shapiro:2001rz}). Such a modification presents a non-propagating additional degree of freedom stemming from the connection -- the torsion tensor (the antisymmetric counterpart of the connection), whose source is the spin density. More generic models, such as metric-affine theories, can take into account dynamical torsion and also other degrees of freedom, such as the non-metricity tensor (see e.g. \cite{Hehl}).

%For example, $f(R)$ gravity theories, which generalize the Einstein–Hilbert action by promoting the Ricci scalar $R$ to an arbitrary function of it, $f(R)$ \cite{Olmo:2011uz}. Yet, }

At the same time, within the gravity context, an important line of investigation is presented by studies of coupling gravity to other fields, especially to the electromagnetic one. In this environment, it is especially important to study new terms which can arise in Einstein-Cartan theory and involve gauge fields. It is still very interesting to study possible generalizations of the Chern-Simons (CS) term, known to play a significant role in electrodynamics in three space-time dimensions, where the photon acquires a nonzero mass without the need for the Higgs mechanism.  Its gravitational analogue term, that is, the gravitational CS term has been introduced in $(2+1)$-dimensional spacetime in \cite{Deser:1982vy},  and in the $3+1$ dimensional spacetime in \cite{Jackiw:2003pm},  where this term breaks the CPT symmetry and, in special cases, the (local) Lorentz symmetry. In \cite{NeeYan}, quantum and classical aspects of the Dirac action in a Riemann-Cartan background were studied using the proper-time method, and they have been shown to imply in the interaction between the Nieh–Yan and Pontryagin topological currents, with the resulting term was shown to be similar to the 4-dimensional gravitational CS term \cite{Jackiw:2003pm}, where the role of the Lorentz-violating (LV) vector played by the torsion. Motivated by \cite{NeeYan}, we propose to generalize such results by including an electromagnetic field minimally coupled to spinors, with the subsequent induction of mixed terms involving gauge and gravity through the proper-time method.  Special attention will be paid to the perturbative generation of the torsion-dependent analogue of the Carroll-Field-Jackiw (CFJ) term, the simplest LV term, defined in \cite{CFJ} and studied in great detail in \cite{JK}, playing in electrodynamics a role similar to the 4-dimensional gravitational CS term; see the discussion in \cite{Jackiw:2003pm}, so that in our case, as in a full analogy with \cite{NeeYan}, this term will arise as a one-loop correction.

This paper is organized as follows. In Section~\ref{sec1}, we start with a review of the coupling between gauge fields and Dirac spinors in a curved spacetime with torsion. In Section~\ref{sec2}, we apply the proper-time method to derive the mixed topological terms. Finally, in Section~\ref{conclu}, we summarize our results.

\section{Dirac-Maxwell fields in gravity with torsion}\label{sec1}
In the Riemann-Cartan geometry, the torsion tensor $T^{\lambda}_{\mu\nu}$ arises as an additional degree of freedom of the connection, which is defined as the antisymmetric piece of the connection $\tilde{\Gamma}^{\lambda}_{\;\mu\nu}-\tilde{\Gamma}^{\lambda}_{\;\nu\mu}=T^{\lambda}_{\;\mu\nu}$. It is a central feature of this approach that distinguishes it from the standard pseudo-Riemannian geometry, which can be recovered by requiring the connection to be symmetric, $\Gamma^{\lambda}_{\;\nu\mu}$. In this section, we review the method that allows us to arrive at the action that involves the spinor and gauge fields in a curved spacetime with torsion. Within this discussion, we follow \cite{NeeYan,Shapiro:2001rz}. 

In complete analogy to \cite{NeeYan}, we start with the Dirac action in $(3+1)$ dimensions defined in a Riemann-Cartan geometry: 
\begin{align}\label{action1}
    S_{D}=\int d^4 x\sqrt{-g} \left[ \frac{i}{2}e^{\mu}_a\left( \bar{\Psi}\gamma^{a}\left(\nabla_\mu  \Psi\right)-\left(\nabla_\mu \bar{\Psi}\right)\gamma^a \Psi\right)-m\bar{\Psi}\Psi\right],
\end{align}
where $\Psi$ is the spinor field, $m$ is its mass and $\gamma^{a}$ are the Dirac matrices. In this scenario, the covariant derivative that includes the spin connection with torsion is given by
\begin{align}\label{derivative1}
    \nabla_\mu \Psi=\partial_\mu \Psi- \frac{i}{2}\tilde{\omega}_{\mu ab} \sigma^{ab}\Psi \;,
\end{align}
 with $\sigma^{ab}=\frac{i}{4}[\gamma^a,\gamma^b]$ and $ [\gamma^a,\gamma^b]=\gamma^a\gamma^b-\gamma^b\gamma^a$ denoting the Dirac matrices commutator. A convenient way of calculations consists in rewriting the covariant derivative in Eq.~\eqref{derivative1}, splitting it into a sum of the torsion-dependent part and the torsion-free one. To do this, we rewrite the spin connection in the form 
\begin{align}
 {\tilde\omega_{\mu}} ^{\;ab}   = \omega_{\mu}^{\;ab} +{K^{ba}}_\mu\;,
 \label{omega}
\end{align}
where $\omega_{\mu}^{\;ab}$ is the Cartan spin connection, defined as
\begin{equation}
    \omega_{cab}=\Omega_{bca}-\Omega_{cab}-\Omega_{acb},
\end{equation}
where $\Omega_{abc}=\frac{1}{2}e^{\phantom{a}\mu}_be^{\phantom{a}\nu}_c(\partial_{\mu}e_{\nu a}-\partial_{\nu}e_{\mu a})$,
and $K^{\mu}_{\phantom{a}\alpha\beta}$ is the so-called contorsion tensor, defined as below
\begin{equation}
   K^{\mu}_{\phantom{a}\alpha\beta}=\frac{1}{2}\left(T^{\mu}_{\phantom{a}\alpha\beta}-T_{\beta\phantom{\mu}\alpha}^{\phantom{b}\mu}-T_{\alpha\phantom{\mu}\beta}^{\phantom{b}\mu}\right),
\end{equation}
and $K^{ab}_{\phantom{ab}\beta}=K^{\mu}_{\phantom{a}\alpha\beta}\,e_{\mu}^{\phantom{a}a}\,e^{\alpha b}$.

 Substituting Eq.\eqref{omega} in  Eq.~\eqref{derivative1}, we get
\begin{align}\label{derivative3}
    \nabla_\mu \Psi=D_\mu \Psi-\frac{i}{2}{K^{ab}}_{\mu}\sigma_{ab}\Psi \;,
\end{align}
with $D_\mu=\partial_\mu-\frac{i}{2}\omega_{\mu ab}\sigma^{ab}$. The above expression allows us to rewrite the action in Eq.~\eqref{action1} as
\begin{align}\label{action2}
  S_{D}=\int d^4 x\sqrt{-g} \bar{\Psi} \left[i\gamma^{\mu} D_{\mu}-m-\frac{1}{8}{K^{ab}}_{\mu}\left(\gamma^{\mu}\sigma_{ab}-\sigma_{ab}\gamma^{\mu}\right) \right]\Psi.
\end{align}
Using the algebraic relation of Dirac matrices $\left\{\gamma^c,\left[\gamma^a,\gamma^b\right]\right\}=4i\varepsilon^{abcd}\gamma_5 \gamma_d$ and the definition of the contorsion tensor, we can obtain the action in Eq.~\eqref{action2} in terms of the axial vector part of the torsion tensor $S_{\mu}=\varepsilon_{\mu\nu\rho\sigma}T^{\nu\rho\sigma}$ (which, within our study, is assumed to be constant; we note that this choice is consistent with the fact that the torsion has no dynamics in the Riemann-Cartan framework), as follows:
 \begin{align}\label{actionsmu}
   S_{D}=\int d^4 x\sqrt{-g} \bar{\Psi} \left[i\gamma^{\mu} D_{\mu}-m+\frac{1}{8}S_{\mu}\gamma_{5}\gamma^{\mu} \right]\Psi\;.
\end{align}

To explore the interaction between fermions and the gauge field $A^{\mu}$ within the Einstein–Cartan framework, we begin by considering the minimally coupled Dirac–Maxwell action in $(3+1)$-dimensional spacetime (see \cite{Belyaev:1997zv}), including dynamical and mass terms for the torsion field, with $S_{\mu\nu}=\partial_{\mu}S_{\nu}-\partial_{\nu}S_{\mu}$ is the analogue of the Maxwell tensor for the torsion: 
\begin{eqnarray}\label{action4}
     S&=&\int d^4 x\sqrt{-g}\left[-\frac{1}{4}F_{\mu\nu}F^{\mu\nu}-\frac{1}{4}S^{\mu\nu}S_{\mu\nu}-ie\epsilon^{\mu\nu\lambda\rho}S_{\mu\nu}F_{\lambda\rho}+\frac{m^2}{2}S^{\mu}S_{\mu}+\right.\nonumber\\&+&\left. \bar{\Psi} \left(i\gamma^{\mu} D_{\mu}-m +\frac{1}{8}S_{\mu}\gamma_5\gamma^{\mu}-e\gamma^{\mu}A_{\mu} \right)\Psi \right] ,
     \end{eqnarray}
or as well in terms of currents,
\begin{eqnarray}
    S&=&S_{M}+\tilde{S}_{D}+\int d^4x\Big[-\frac{1}{4}S^{\mu\nu}S_{\mu\nu}-ie\epsilon^{\mu\nu\lambda\rho}S_{\mu\nu}F_{\lambda\rho}+\frac{m^2}{2}S^{\mu}S_{\mu}\Big]-\nonumber\\&-&\int d^4 x\sqrt{-g}\left[eJ^{\mu}A_{\mu}+\frac{1}{8}S_{\mu}J^{\mu}_5 \right].
\end{eqnarray}
This form of the gauge-torsion action is motivated by the one-loop renormalizability reasons. We note, however, that within our calculations, the torsion is assumed to be constant (background field, hence all terms involving $S_{\mu\nu}$ vanish in our case. Additionally, we note that the CFJ-like gauge-torsion term, which will be generated in the next section, is finite and then does not need to be introduced from the very beginning, unlike torsion-dependent terms in (\ref{action4}).
In the above equation, $S_{M}$ and $\tilde{S}_{D}$ are the Maxwell action and pure Dirac action without torsion, respectively. Note that this action involves two currents, first, the usual Dirac current $J^{\mu}=\bar{\Psi}\gamma^{\mu}\Psi$ coupled to the gauge field $A_{\mu}$, like the standard QED, second, the axial fermion current $J^\mu_{5} = \bar{\Psi} \gamma^\mu \gamma^5 \Psi$ coupled to the axial vector $S_{\mu}$ absent in the usual torsionless Dirac action.

We adopt the background field method, which consists of considering the axial vector piece of the torsion as a background field. This choice preserves gauge invariance and the original form of the $U(1)$ transformation, which would otherwise be compromised due to the known incompatibility between torsion and gauge symmetry, as discussed in ~\cite{Kibble:1961ba, Shapiro:2001rz}. The possibility of tying the two theories together has been discussed, for example, in \cite{Nieh:2017zxr, Hojman:1978yz, Novello:1976tx}. In particular, Ref.~\cite{Nieh:2017zxr} presents a construction of a symmetric connection that maintains both gauge and scale invariance in a massless Maxwell–Dirac system coupled to torsion. Nevertheless, since the photon field does not contribute to the spin current and hence does not act as a source of torsion, our choice to retain the standard gauge structure remains justified. 
 
Actually, since the gauge field does not display the properties of the spinor, it does not act as a direct source of torsion, but it can still be indirectly affected by the quantum fluctuations of the spinor in the presence of torsion; mixed gauge–torsion couplings can arise as a quantum correction  \cite{Ojima,ourptime}. In this context, the functional approach offers a method for computing these induced effects. Next, the Schwinger-deWitt proper-time method will be applied in order to investigate effective couplings between torsion and gauge fields and extensions of models with a gravitational Chern-Simons term.

\section{Proper-time method in Maxwell-Einstein-Cartan QED}\label{sec2}
Our task in this section is to get terms that couple torsion with the gauge field. For it, we start with the one-loop effective action of gauge and gravitational fields $\Gamma^{(1)}[A_{\mu},g_{\nu\lambda}]$ generalizing the analogous expressions considered in \cite{Ojima} and \cite{ourptime}:
\bea
e^{i\Gamma^{(1)}[A_{\mu},g_{\nu\lambda}]}=\int D\Psi D\bar{\Psi}e^{iS},
\eea
where $S$ is the classical action of our theory given by (\ref{action4}), therefore, the corresponding one-loop effective action can be written as
\bea
\label{efa}
\Gamma^{(1)}=-i{\rm Tr}\ln (-i\Ds-m-\slashed{S}\gamma_5-e\slashed{A}).
\eea
Here, we use the standard slashed notation to represent the contraction involving the Dirac matrices $\gamma^{\mu}$, where the slashed terms are given by $\slashed{S}=S_\mu\gamma^\mu$, for the gauge field $\slashed{A}=\gamma^{\mu}A_{\mu}$ and $\os=\gamma^\mu \omega_{\mu ab}\sigma^{ab}$ (hence, $i\ds=i\Ds-\os$  is field independent). In this expression, the factor $\frac{1}{8}$ accompanying the torsion in (\ref{action4}) is absorbed into the definition of this vector; also, some redefinitions of signs are carried out. Adding to the effective action (\ref{efa}) the term given by $i{\rm Tr}\ln (i\Ds-\os-m)$ which does not depend neither on $A_{\mu}$ nor on the torsion, and using the identity $\det(AB)=\det A\det B$, which implies $\ln \det(AB)=\ln \det A+\ln \det B$, together with $\ln \det A={\rm Tr}\ln A$, we have:
\bea
\label{efa2}
\Gamma^{(1)}=-i{\rm Tr}\ln [(-i\Ds-m-\slashed{S}\gamma_5-e\As)(i\Ds-\os-m)].
\eea
Then, it is known that $\Ds\Ds=D^2+\frac{R}{4}$ since the commutator of covariant derivatives is the curvature tensor.
Hence, opening the product of our operators in the equation above, we have
\bea
\label{efa3}
\Gamma^{(1)}&=&-i{\rm Tr}\ln [D^2+m^2+\frac{R}{4}+m\os-ie\As\Ds+i\slashed{S}\Ds\gamma_5+i\Ds\os+e\As\os-\slashed{S}\os\gamma_5+em\As +\nonumber\\&+&
m\slashed{S}\gamma_5].
\eea
Here we apply the Schwinger-DeWitt proper time description (see \cite{BV} for a review):
\bea
\label{efa4}
\Gamma^{(1)}&=&-i{\rm Tr}\int\frac{ds}{s}\int d^4x \exp\Big(is[D^2+m^2+\frac{R}{4}+m\os-ie\As\Ds+i\Ss\Ds\gamma_5+i\Ds\os+\nonumber\\ &+& e\As\os-\Ss\os\gamma_5+em\As +m\Ss\gamma_5]\Big)i\delta^4(x-x')|_{x=x'},
\eea
where $s$ is the Schwinger proper time parameter. We note that within our framework $S_{\mu}$ is assumed to be constant.

Now we can expand the exponential in power series in fields and use the cyclic property of the trace:
\bea
\label{efa5}
\Gamma^{(1)}&=&-i{\rm Tr}\int\frac{ds}{s}\int d^4x \Big(1+is[m\os-ie\As\Ds+i\Ss\Ds\gamma_5+i\Ds\os+e\As\os-\Ss\os\gamma_5+em\As +m\Ss\gamma_5]+\nonumber\\
&+&\frac{1}{2!}(is)^2[m\os-ie\As\Ds+i\Ss\Ds\gamma_5+i\Ds\os+e\As\os-\Ss\os\gamma_5+em\As +m\Ss\gamma_5]^2+\nonumber\\
&+&\frac{1}{3!}(is)^3[m\os-ie\As\Ds+i\Ss\Ds\gamma_5+i\Ds\os+e\As\os-\Ss\os\gamma_5+em\As +m\Ss\gamma_5]^3+
\ldots
\Big)
\times\nonumber\\&\times&
\exp\left(is[D^2+m^2+\frac{R}{4}]\right)\delta^4(x-x')|_{x=x'}.
\eea
Then we use the formulas from \cite{Ojima}: $\delta^4(x-x')=\int\frac{d^4k}{(2\pi)^4}e^{ik^aD_a\sigma(x,x)}$, $\lim\limits_{x\to x'}\sigma(x,x')=0$, $\lim\limits_{x\to x'}D_m\sigma(x,x')=0$, $\lim\limits_{x\to x'}D_mD_n\sigma(x,x)=g_{mn}$, etc. Also, within our approximation, we assume the curvature to be small, so we concentrate on terms of zero order in the scalar curvature. At the same time, we want to consider the contributions that are non-trivially dependent on the torsion, which is known to be small and non-dynamical \cite{Shapiro:2001rz}, hence we keep only the terms of the first order in the torsion.

This calculation naturally generalizes the one carried out in \cite{ourptime, NeeYan} by adding new terms. Therefore, repeating the calculations performed in \cite{NeeYan}, we can show that the result for the Nieh-Yan term, that is, the analogue of the gravitational Chern-Simons term, where the axial LV vector is presented by the torsion, reproduces the result given in \cite{NeeYan}, that is,
\bea
\label{ny}
S_{ny}=\frac{1}{216\pi^2}\int d^4x\sqrt{-g}\epsilon^{\mu\nu\lambda\rho}S_{\mu}(\omega_{\nu ab}\partial_{\lambda}\omega_{\rho}^{\phantom{\rho}ab}+\frac{2}{3}\omega_{\nu ab}\omega_{\lambda}^{\phantom{\lambda}bc}\omega_{\rho c}^{\phantom{\rho c}a}).
\eea
Then, let us consider the contributions dependent on the gauge field, whose study is our main aim in this paper. We concentrate on the contributions quadratic in this field and linear in torsion, so, their form is expected to be similar to the CFJ term. The relevant gauge-torsion term carries $(is)^3$ factor (all lower-order terms evidently will be either at most of the first order in $A_m$ or torsion independent and hence are irrelevant within our study):
\bea
\label{eafinal}
\Gamma^{(1)}&=&-i{\rm Tr}\int\frac{ds}{s}\int d^4x \Big(
\frac{1}{3!}(is)^3[-ie\As\Ds+i\Ss\Ds\gamma_5+em\As +m\Ss\gamma_5]^3
\Big)
\times\nonumber\\&\times&
\exp\left(is[D^2+m^2]\right)\delta^4(x-x')|_{x=x'}+\ldots,
\eea
where the dots are for irrelevant terms depending on $A_m$ and connection at the same time (actually, being considered independently, they are incompatible with gauge symmetry since isolated terms of the first order in the connection are not covariant. We note that since the curvature tensor is constructed on the base of the connection, while the torsion is independent of the curvature, such terms probably can be combined with other connection-dependent terms forming the objects proportional to the curvature tensor and higher-order tensors constructed on its base, but they will contribute only to higher orders of the derivative expansion of the one-loop effective action). The term given by (\ref{eafinal}) is actually finite since the corresponding integral over $s$ converges. Keeping here only the relevant terms (first orders in derivatives and torsion and the second order in $A_{\mu}$), we find
\bea
\label{eafinal2}
\Gamma^{(1)}&=&-i{\rm Tr}\int\frac{ds}{s}\int d^4x \Big(
(is)^3[-ie^2m^2 \As\ds\As\Ss\gamma_5]\Big)
\times\nonumber\\&\times&
\exp\left(is[D^2+m^2]\right)\delta^4(x-x')|_{x=x'}+\ldots,
\eea
where we consider the case of zero connection; at the same time, by covariance reasons, the connection generates only higher-order, curvature dependent contributions. It remains to calculate the trace: we use the fact that $e^{isD^2}\delta^4(x-x')= \frac{1}{16\pi^2s^2}$, and ${\rm tr}(\gamma^{\mu}\gamma^{\nu}\gamma^{\lambda}\gamma^{\rho}\gamma_{5})=4i\epsilon^{\mu\nu\lambda\rho}$. So, we rest with
\bea
\Gamma^{(1)}=-im^2\int\frac{ds}{s}\frac{1}{4\pi^2s^2}s^3e^{-sm^2}\epsilon^{\mu\nu\lambda\rho}A_{\mu}\partial_{\nu}A_{\lambda}S_{\rho}=\frac{1}{4\pi^2}\epsilon^{\mu\nu\lambda\rho}A_{\mu}\partial_{\nu}A_{\lambda}S_{\rho}.
\eea
So, we successfully generated the Carroll-Field-Jackiw-torsion (CFJ-torsion) term.  We note that, while this term is formally not gauge independent, its contribution to the effective action is gauge-independent for the constant vector $S_{\mu}$, just the same situation is known to take place for the standard CFJ term, see \cite{CFJ}. Note, however, that in addition to CFJ-torsion and NY terms, in the non-zero curvature case there can exist the following mixed term 
\begin{equation}  \Gamma^{(1)}_{mixed}\propto\epsilon^{\mu\nu\lambda\rho}R_{[\mu\nu]}A_{\lambda}S_{\rho},
\end{equation}
where $R_{[\mu\nu]}$ is the antisymmetric part of the Ricci tensor, which in general is non-trivial in the Riemann-Cartan geometry. 
%\begin{equation}
%    R_{[ab]}...
%\end{equation}
This term will evidently be finite; however, its nature requires a more profound discussion, which we will present elsewhere.

The Nieh-Yan and CFJ-torsion terms are the lower (linear in torsion) contributions to the effective action allowed by the covariance reasons. Other contributions will involve higher derivatives (including curvature-dependent terms). 

Therefore, we conclude that we have succeeded in calculating the lower quantum contributions in Einstein-Cartan gravity with the gauge field, both in the gravitational and gauge sectors.

\section{Conclusion}\label{conclu}
We considered the theory involving couplings of the spinor field to Einstein-Cartan gravity and the gauge field at the same time. For this theory, we calculated the lower contributions to the one-loop effective action. We explicitly demonstrated that the result depending on the gauge field reproduces the CFJ form. It is clear that this calculation can easily be generalized for the non-Abelian case, following the lines of \cite{ourptime}, with the same numerical coefficient as in this study.

It is important to emphasize that, similarly to \cite{NeeYan}, our result allows us to suggest that one of the possible mechanisms for the Lorentz symmetry breaking, for the axial LV vector, consists in the presence of some primordial non-zero torsion which naturally introduces a space-time anisotropy and hence breaks the Lorentz symmetry. In this context, it is worth mentioning that a possible relation between Lorentz symmetry breaking and torsion has also been indicated in \cite{Alexander:2008wi}. 

In principle, one can also consider other contributions generated by our one-loop effective action (\ref{efa3}). In particular, in the lower orders of its expansion, one can expect the arising of the term proportional to $\epsilon^{\mu\nu\lambda\rho}S_{\mu}A_{\nu}R_{\lambda\rho}$ (we note that the Ricci tensor is in general not symmetric within the non-Riemannian geometry). A detailed analysis of this term will be performed elsewhere.

We note that the CFJ-torsion term can be naturally embedded within the possible generalization of the LV SME presented in \cite{Kostelecky:2003fs,Kostelecky:2020hbb}, where, in addition to the detailed discussions of the LV SME within the Riemannian geometry context, some discussions of possible LV terms in the presence of torsion are given. A detailed classification of such terms is still to be done. We hope to study other torsion-dependent contributions to effective action in forthcoming papers.

%%%%%%%%%%%%%%%%%%%%%%%%%%%%%%%%%%%%%%%

{\acknowledgments} The authors would like to thank the Conselho Nacional de Desenvolvimento Cient\'{\i}fico e Tecnol\'{o}gico (CNPq) for financial support. P. J. Porf\'{\i}rio would like to acknowledge the Brazilian agency CNPq, grant No. 307628/2022-1. M. Paganelly acknowledges grant support from CNPq 152508/2024-4. The work by A. Yu. Petrov has been partially supported by the CNPq project No. 303777/2023-0.

\end{document}